\documentclass[a4paper,11pt,fleqn]{article}

\usepackage[centertags]{amsmath}
\usepackage{amssymb,bm}
\usepackage{amsfonts}

\usepackage{graphicx}
\usepackage{color}
\usepackage{placeins}
\usepackage[square]{natbib}

\newcommand{\dt}{\delta t}

\newcommand{\DT}{\Delta T}

\newcommand{\E}[1]{E\!\left[ #1 \right]}

\newcommand{\cdf}{\operatorname{cdf}}

\begin{document}


\begin{center}
 { \bf \huge On the short term stability \\[1ex] of financial ARCH price processes}

\vspace{5ex}
{\bf\large
	Gilles Zumbach
}\\[2ex]
\parbox{0.4\textwidth}{\renewcommand{\baselinestretch}{1.0}\normalsize
Edgelab\\
Avenue de la Rasude 5\\
1006 Lausanne\\
Switzerland
}
\\[3ex]
\parbox{0.4\textwidth}{\renewcommand{\baselinestretch}{1.0}\normalsize
gzumbach@edgelab.ch\\
gilles.zumbach@bluewin.ch\\

\vspace*{4ex}
\today
}


\begin{abstract}
For many financial applications, it is important to have reliable and tractable models for the behavior of assets and indexes, for example in risk evaluation.
A successful approach is based on ARCH processes, which strike the right balance between statistical properties and ease of computation.
This study focuses on quadratic ARCH processes and the theoretical conditions to have a stable long-term behavior.
In particular, the weights for the variance estimators should sum to 1, and the variance of the innovations should be 1.
Using historical data, the realized empirical innovations can be computed, and their statistical properties assessed.
Using samples of 3 to 5 decades, the variance of the empirical innovations are always significantly above 1, for a sample of stock indexes, commodity indexes and FX rates.
This departure points to a short term instability, or to a fast adaptability due to changing conditions.
Another theoretical condition on the innovations is to have a zero mean.
This condition is also investigated empirically, with some time series showing significant departure from zero.

\end{abstract}
\end{center}
\vspace{6ex}
Keywords: \parbox[t]{0.8\textwidth}{ARCH processes, market stability, innovations, drift significance}\\
JEL: ~~C22, C62, C52

\newpage

\section{Introduction}

This paper is about daily time series and the corresponding mesoscopic processes.
An important topic is to write a process, or a family of processes, that describes the statistical behavior of market data along the time direction.
Possible practical applications of these processes are in risk evaluation, portfolio allocations, and in option pricing.
The base model is a normal random walk with constant volatility, postulated by \citep{Bachelier.1900} in his thesis.
Since, the accumulation of digital data and the computational power of modern computers allow the evaluation of detailed empirical statistics, and to test finer models.
A key challenge is to describe the dynamic aspect of the volatility.
With that respect, the ARCH family of processes has been very successful.
The literature on volatility processes and/or forecast is very large, for example see \citep{Engle.1982, EngleBollerslev.1986} for the original references, \citep{AndersenBollerslev.1998} for the volatility forecasts induced by a GARCH(1,1) process, \citep{HansenLunde.2005} for general ARCH forecasts, or \citep{AndersenBollerslevChristoffersenDiebold.2006, Poon.2005, Zumbach.book} for broad reviews.
The key idea of the ARCH model is to have a volatility that depend on the past change of prices, introducing a feed-back loop between the price evolution and the volatility.
In this class of processes, the Long-Memory ARCH (LM-ARCH) processes \citep{Zumbach.LongMemory} describes correctly the clustering of the volatility, the decay of the lagged correlations of the volatility, and the quantitative measures of the time reversal non-invariance \citep{Zumbach.TimeReversalInvariance}.
Importantly for wide risk evaluation, they depend on a few parameters with a very weak dependency on each time series.
A related process with long memory has been proposed by \citep{Corsi.2009}.

Regardless of the ARCH process details, we consider the class of quadratic processes where the variance for the next time step is given by a weighted sum of the past squared returns, with decaying weights as the returns are further away in the past.
For the process to be consistent, the weights should sum to one.
In a process, our ignorance about the future events are represented by a random variable, typically called the innovations.
Again for the process consistency, the variance of this random variable should be one.
Both conditions are required so that the variance of the price changes has a non-singular asymptotic distribution, and does not decrease to zero or increase without bounds.

For a given decay shape of the memory weights, two versions of an ARCH process can be written.
In the linear form, the variance is a weighted sum of past squared returns.
In the affine form, a constant is added to the weighted sum that fixes the asymptotic mean variance of the process.
The above constraints on the weights and innovation variance are needed so that the mean volatility parameter is equal to the long term mean volatility of the process.

The decay shape for the weights in the volatility evaluation is otherwise free and several shapes are commonly used, like Rectangular Moving Average (RMA), Exponential Moving Average (EMA), or decays that are slower than exponential as in the Long Memory (LM) version of the ARCH process.
But all shapes should obey the above constraint that the sum of the weights is one.

Turning to empirical data, the past prices allow to compute the historical daily returns.
For a given quadratic ARCH volatility process, namely for a particular decay shape for the weights, the volatility can be computed.
With the time series of volatilities and realized returns, the process equation can be reverted to compute the realized innovations.
In principle, these realized innovations should obey the consistency constraints of a process, in particular their variances should be one.
Several studies have been done along this idea, see e.g. \citep{AndersenBollerslevDieboldLabys.2000, Zumbach.RM2006_fullReport}, with the main purpose is to study the probability distribution for the innovations.
The focus for this paper is on the standard deviation of the innovations.

On empirical data, the standard deviation of the innovations is slightly larger than one, of the order of 3 to 4\% larger.
The deviations are small but significant, and consistent for many time series (stock indices, commodity indices, FX).
The LM-ARCH weights give the smallest deviations quoted above, other shapes give larger deviations from 1.
To summarize, no time series is compatible with an ARCH model using innovations with unit variances.

This deviation from the theoretical criterion points to a slight instability of the empirical data, and of the underlying trading mechanism.
Since systematically larger than one, the volatility should increase exponentially fast.
This is obviously not the case, hence another mid-term mechanism should provide for the long term stability.
For example, an Ornstein-Uhlenbeck term (in the volatility or in the price equation) can provide for a medium to long term stability.
Since the deviations from 1 are small, a small recoil term is sufficient to bring a long term stability.

It is not unreasonable to have a short term instability.
This means that the short-term trading is very reactive to external information or to endogeneous perturbations.
This reactivity can be seen as a quality since the system is adapting very quickly to a changing environment.
The trading activity can be a bit over-reacting, with the advantage to reach quickly a new equilibrium.
At a medium to long term scales, some fundamental criterion bring the system back, leading to the observed long term stability.


Another theoretical condition sets on the innovations is to have a zero mean, while the possible drift for the price is contained in a mean term in the ARCH equation for the returns.
This mean return term is typically small and difficult to estimate, because the volatility and its dynamics dominates the price processes.
In particular, any mean term equation is difficult to validate out-of-sample.
As a simple solution, we set it to zero in the empirical study.
Consequently, the possible mean return is reported on the mean empirical innovations.
The subsequent question is if the empirical innovations are compatible with a zero mean.
It is the case for most FX rates and commodity indexes, while it is not compatible with a zero mean for most stock indices.
For this last category of time series, this is the expected long term results, namely stock indices drift upward.
Yet, some stock indices are compatible with a zero mean for the innovations, indicating that they did not increase significantly over the last few decades.

The rest of the paper is organized as follows.
The next section presents the core material on quadratic ARCH processes and on the consistency conditions.
Sec.~\ref{sec:statistics} introduces the statistical material using the central limit theorem, and in particular the computation of the $p$-values when the random variables are distributed with a Student distribution.
The empirical results for a LM-ARCH volatility are given in Sec.~\ref{sec:empiricalMean} for the means, and Sec.~\ref{sec:empiricalVariance} for the variances.
Empirical results for other volatility kernels are discussed in Sec.~\ref{sec:otherVolatilityEstimators}, some further investigations reported in Sec.~\ref{sec:furtherInvestigations}, before the conclusions.

\section{Relevant background on ARCH models}
\label{sec:archTheory}
A wide array of ARCH processes have been proposed to model financial time series.
We will consider in this paper the family of processes where the variance is a linear or affine function of past squared returns.
The ARCH process is essentially a location/size/shape decomposition of the random returns $r(t)$, namely
\begin{equation}
\label{eq:locSizeShape}
r(t+\dt) = \tilde{\mu}(t) + \widetilde{\sigma}(t)\, \epsilon(t+\dt) \hspace{3em} \epsilon ~~\text{i.i.d with distribution}~~ p(\epsilon).
\end{equation}
The terms $\tilde{\mu}(t)$ and $\widetilde{\sigma}(t)$ fix the location and size for the next random return $r(t+\dt)$.
They are function of the information set up to $t$, and give a one-step forecast for the mean and standard deviation of the random variable $r(t+\dt)$.
The random variable $\epsilon(t+\dt)$ is called the innovation and has a fixed distribution $p(\epsilon)$ that is time independent.
The distribution $p$ has a zero mean and unit variance, but is otherwise unspecified.
With this structure, all the time dependencies on the history of $r$ are included in the predictable parts  $\tilde{\mu}(t)$ and $\widetilde{\sigma}(t)$, while $p(\epsilon)$ has no dependency on the history or on time.

Using the properties of $p(\epsilon)$, the mean and variance of $r$ are
\begin{subequations}
	\begin{align}
	\E{r(t+\dt)} & = \tilde{\mu}(t) \\
	\E{\big(r(t+\dt) - \tilde{\mu}(t)\big)^2} & = \widetilde{\sigma^2}(t)
	\end{align}
\end{subequations}
We are making the convenient abuse of notation
\begin{equation}
   \widetilde{\sigma} = \sqrt{\widetilde{\sigma^2}}.
\end{equation}
In this paper, we consider the class of forecasts where $\widetilde{\sigma^2}$ is a function of the past squared returns, possibly with an additive constant for the affine processes.

At a one-day scale, the financial time series are dominated by the random increment $\epsilon$ and by the volatility dynamics $\widetilde{\sigma}$, while the mean term $\tilde{\mu}$ is small and very difficult to estimate.
Consequently, we will take $\tilde{\mu} = 0$ afterward.

For a \emph{linear} model, the variance is a linear function of the past squared returns
\begin{subequations}
\begin{align}
	\widetilde{\sigma^2}_\text{lin}(t) & = \sum_{l=0}^{l_\text{max}} w(l)\,r^2(t-l) \\
	\sum_{l=0}^{l_\text{max}} w(l) & = 1 \hspace{3em}w(l) \geq 0 \label{eq:weightsNormalizationLin}
\end{align}
\end{subequations}
where $t$ is the current time, $l$ is the lag from the current time, and $r(t-l)$ is the one day return at the time $t-l$.
The weights $w(l)$ specifies the importance of the past events on today volatility forecast.
They are typically decreasing for increasing lag, quantifying the progressive decay of the past information on the forthcoming events.
In practice, they are defined on a bounded support for numerical evaluation.
Some standard shapes are the constant weights $w(l) = 1/n$ in a window of length $n$ (Rectangular Moving Average), and the exponentially decaying weights $w(l) = \exp(-l/\tau)$ (Exponential Moving Average with $\mu = \exp(-1/\tau)$).
An intermediary shape is given by the long memory weights proposed in \citep{Zumbach.LongMemory}, where the weights decay as a logarithm of the lag.
This last shape gives lagged correlations for the volatility that are very close to the empirical ones.

For a linear ARCH model, the equation \ref{eq:weightsNormalizationLin} is a consistency condition for the volatility, which ensure that $\E{r^2(t)}$ is given by the initial conditions of the process.
If the sum of the weights is smaller (larger) than one, the expected variance goes to zero (infinity) exponentially fast for increasing time.
Essentially, the conditions $\sum w(l) = 1$ and $\E{\epsilon^2} = 1$ are consistency conditions so that the volatility stays around the initial values.
If these conditions are not fulfilled, the process goes to a singular state, either to a fixed price (zero volatility) or to in infinite volatility.

For an \emph{affine} model, the variance can be written as a convex combination between the mean variance and a linear ARCH model
\begin{align}
	\widetilde{\sigma^2}_\text{aff}(t) & = w_\infty \sigma_\infty^2
		+ (1 - w_\infty)\widetilde{\sigma^2}_\text{lin}(t) \label{eq:AffVarianceFromLinVariance} \\
	& = w_\infty \sigma_\infty^2 + \sum_{l=0} (1 - w_\infty) w(l)\,r^2(t-l) \nonumber
\end{align}
where the parameter $0 \le w_\infty \leq 1$ controls the balance between the mean volatility $\sigma_\infty$ and the auto-regressive part \citep{Zumbach.LongMemory}.
In the second form, the sum of all the coefficients is 1.
Again, this is a consistency condition to ensure that $\E{r^2(t)}$ converges to $\sigma_\infty^2$ for increasing time, namely that $\sigma_\infty$ is the parameter that fixes the mean volatility.

The familiar GARCH(1,1) process can be cast in the form~\eqref{eq:AffVarianceFromLinVariance}, with the linear part $\widetilde{\sigma^2}_\text{lin}$ being an EMA (see \citep{Zumbach.LongMemory} or \citep{Zumbach.book} for the mapping between equations and parameters).
In the present form, the parameters for GARCH(1,1) can be easily interpreted.
The dynamic volatility part is measured by an EMA, with a characteristic time given by the exponential decay coefficient.
The long term asymptotic volatility is $\sigma_\infty$, and the convex combination between both variances is controlled by $w_\infty$.
Essentially, the parameter $w_\infty$ controls the conditional heteroskedasticity, the other parameters control the mean volatility and the dynamical measure of recent past volatility.

When estimated on empirical data, the coefficient $w_\infty$ is always small, of the order of 2 to 10\%.
This value depends on the linear part, more precisely on the decay of the weights $w(l)$.
With a slow decay (i.e. slower than exponential), the parameter $w_\infty$ becomes smaller, in the 2 to 5\% range.
This can be understood as the long term weights inducing a mean volatility measured on the past sample, hence reducing the usefulness of a mean volatility parameter in the equation.

The small value for $w_\infty$  means that the financial markets operate close to the linear regime, where the volatility is purely auto-regressive.
The stabilization provided by $\sigma_\infty$ is weak, leading to large volatility excursions (crises).
Hence, it is a good approximation for the empirical studies to use a linear model instead of an affine one, removing 2 parameters ($w_\infty$ and $\sigma_\infty$).
The most annoying parameter is $\sigma_\infty$, since time series dependent.
Moreover, its main purpose is to provide sane long term statistical properties (see \citep{Nelson.1990} for a detailed analysis for the GARCH(1,1) process), that are irrelevant at short time scale.
Hence, it is simpler to remove these 2 parameters and to work with a minimal process.

With historical data, Eq.~\ref{eq:locSizeShape} can be reverted to obtain the realized innovations
\begin{equation}
\label{def:historicalInnovation}
   \epsilon(t+\dt) = \frac{r(t+\dt) - \tilde{\mu}(t)}{\widetilde{\sigma}(t)}.
\end{equation}
The time series of innovations is the basis for the subsequent analysis.

\section{Statistics on the innovations and p-values}
\label{sec:statistics}
In a process, the innovations $\epsilon$ have a distribution $p(\epsilon)$.
The standard choices for $p$ are a normal distribution and a Student distribution (scaled to have a unit variance).
We will use these two distributions to compute the $p$-values of the empirical mean and variance, using the central limit theorem.

The base quantities for the statistical test are the empirical mean and variance of $\epsilon$
\begin{subequations}
\begin{align}
	\widehat{\mu} & = \frac{1}{n}\,\sum_t \epsilon(t) \\
    \widehat{\sigma^2} & = \frac{1}{n-1}\,\sum_t \left(\epsilon(t) - \hat{\mu}\right)^2 \label{eq:empiricalVariance}
\end{align}
\end{subequations}
with $n$ points in the sum.
The benchmark is to assume a normal distribution or a scaled Student distribution.
For both distributions, the mean of $\epsilon$ is 0 and the variance is 1.
The mean of $\epsilon^2$ is 1, and the variance of $\epsilon^2$ is
\begin{subequations}
	\label{eq:y_variance}
\begin{align}
	\text{var}(\epsilon^2) & = 2 &&  \text{Normal distribution}\\
    \text{var}(\epsilon^2) & = 2\, \frac{\nu-1}{\nu - 4}  \hspace{3em}\nu > 4\hspace{2em} && \text{Student distribution}  \label{eq:y_variance.Student}.
\end{align}
\end{subequations}
These relations can be derived as follows.
For the Student distribution, the kurtosis is
\begin{equation*}
\operatorname{kurt} = \E{\left(\frac{\epsilon - \mu}{\sigma}\right)^4} = 3 + \frac{6}{\nu - 4}
\end{equation*}
which is equal to $\E{\epsilon^4}$ for the scaled Student distribution.
The variance of $\epsilon^2$ for the scaled Student distribution is
\begin{equation*}
  \text{var}(\epsilon^2) = \E{\epsilon^4} - \E{\epsilon^2}^2 = 3 + \frac{6}{\nu - 4} - 1
\end{equation*}
leading to Eq.~\ref{eq:y_variance.Student}.
The result for the normal distributon can be derived similarly, or by taking the large $\nu$ limit of the Student relation.

The central limit theorem can now be used to find the distribution for $\widehat{\mu}$ and $\widehat{\sigma^2}$.
We introduce a variable $z$, scaled so has to have a standard normal distribution.
For the sample mean, in the large $n$ limit, the quantity
\begin{equation}
   z = z(\widehat{\mu}) = \frac{\widehat{\mu} - 0}{\sqrt{1/n}}
\end{equation}
has a normal distribution with zero mean and a unit variance.
For an empirical value $\widehat{\mu}$, the probability of a value as large or larger than $|\widehat{\mu}|$ (two sided test) is
\begin{eqnarray}
    p = 2(1 - \cdf(z(|\widehat{\mu}|)))
\end{eqnarray}
where $\cdf$ is the cumulative distribution function of the normal distribution.

For the variance, the quantity
\begin{equation}
z = z\left(\widehat{\sigma^2}\right) = \frac{\widehat{\sigma^2} - 1}{\sqrt{\text{var}(\epsilon^2)/n}}
\end{equation}
has a normal distribution with zero mean and a unit variance.
For an empirical value $\widehat{\sigma^2}$, the probability of a value as large or larger than $\widehat{\sigma^2}$ (one sided test) is
\begin{eqnarray}
p = 1 - \cdf(z)
\end{eqnarray}
This $p$-value can be computed assuming a normal or Student distribution for $\epsilon$ using Eq.~\ref{eq:y_variance}.

\section{Empirical results for the means}
\label{sec:empiricalMean}
The statistics $\widehat{\mu}$ and $\widehat{\sigma^2}$ with their p-values have been computed for a set of 24 time series with a long time span.
Major stock indexes, commodity indexes and FX are included in the empirical sample.
The results are given in table~\ref{table:LMARCH} for the volatility evaluated with the LM-ARCH weights.
The parameters for the volatility estimator are fixed according to \citep{Zumbach.RM2006_fullReport}, namely they have not been optimized on each time series as done in most studies.
These results are analyzed first, other volatility forecasts discussed shortly in section \ref{sec:otherVolatilityEstimators}.
\begin{table}
\begin{tabular}{|lrr|rr|rrr|}
	\hline
 &   &  & \multicolumn{2}{|c|}{mean}  &  \multicolumn{3}{|c|}{variance}  \\
 &   &  &  &  &   &  \multicolumn{2}{c|}{p-value}  \\
name &  startDate &  length &  mean &  p-value &  variance &  St 6 &  St 5 \\
\hline
CAC 40                              & 1990-03-15 & 7630 & 0.013 & 0.248 & 1.049 & 0.029 & 0.066 \\
DAX                                 & 1962-09-10 & 14494 & 0.027 & 0.001 & 1.054 & 0.002 & 0.011 \\
Dow Jones Industrial Average        & 1962-09-07 & 14546 & 0.035 & 0.000 & 1.060 & 0.001 & 0.005 \\
EuroStoxx                           & 1989-09-06 & 7859 & 0.019 & 0.091 & 1.067 & 0.004 & 0.018 \\
FTSE 100                            & 1986-09-05 & 8505 & 0.021 & 0.051 & 1.051 & 0.018 & 0.048 \\
Hong Kong Hang Seng Index           & 1969-11-24 & 12373 & 0.047 & 0.000 & 1.069 & 0.000 & 0.003 \\
IBEX 35                             & 1989-09-11 & 7721 & 0.014 & 0.232 & 1.086 & 0.000 & 0.004 \\
NASDAQ 100                          & 1987-10-12 & 8199 & 0.049 & 0.000 & 1.051 & 0.019 & 0.050 \\
Nikkei 225                          & 1972-09-11 & 11718 & 0.022 & 0.019 & 1.078 & 0.000 & 0.001 \\
OMX STOCKHOLM 30 INDEX              & 1989-08-24 & 7697 & 0.028 & 0.013 & 1.063 & 0.007 & 0.026 \\
\hline
S\&P GSCI Agriculture                & 1972-09-06 & 12014 & 0.011 & 0.218 & 1.065 & 0.001 & 0.006 \\
S\&P GSCI Coffee                     & 1983-09-14 & 9241 & 0.014 & 0.173 & 1.082 & 0.000 & 0.003 \\
S\&P GSCI Cotton                     & 1979-09-14 & 10250 & 0.010 & 0.288 & 1.089 & 0.000 & 0.001 \\
S\&P GSCI Energy                     & 1985-09-13 & 8736 & 0.012 & 0.272 & 1.072 & 0.001 & 0.009 \\
S\&P GSCI Gold                       & 1980-09-12 & 9999 & 0.017 & 0.082 & 1.098 & 0.000 & 0.000 \\
S\&P GSCI Grains                     & 1972-09-06 & 12015 & 0.012 & 0.187 & 1.066 & 0.001 & 0.005 \\
S\&P GSCI Light Energy               & 1972-09-08 & 12013 & 0.014 & 0.125 & 1.075 & 0.000 & 0.002 \\
S\&P GSCI Non Precious Metals        & 1976-09-14 & 11004 & 0.010 & 0.289 & 1.072 & 0.000 & 0.004 \\
S\&P GSCI Soybeans                   & 1972-09-06 & 12015 & 0.025 & 0.007 & 1.051 & 0.006 & 0.023 \\
\hline
CAD-USD                             & 1973-09-10 & 12164 & -0.014 & 0.126 & 1.061 & 0.001 & 0.008 \\
CHF-USD                             & 1973-09-10 & 12164 & 0.022 & 0.016 & 1.175 & 0.000 & 0.000 \\
EUR-USD                             & 1977-09-08 & 10950 & 0.002 & 0.850 & 1.072 & 0.000 & 0.004 \\
GBP-USD                             & 1973-09-10 & 12158 & -0.005 & 0.583 & 1.082 & 0.000 & 0.001 \\
JPY-USD                             & 1973-09-10 & 12164 & 0.015 & 0.098 & 1.063 & 0.001 & 0.007 \\
	\hline
\end{tabular}
\caption{Statistics for $\widehat{\mu}$ and $\widehat{\sigma^2}$ with their p-values.
	The volatility is computed using LM-ARCH weights.
	The column with header ``St 6'' contains the $p$-value for the variance assuming a Student distribution for the innovations with 6 degrees of freedom (and similarly for the column ``St 5''). }
\label{table:LMARCH}
\end{table}

The mean of $\epsilon$ is significantly departing from zero for roughly two third of the stock indexes, for the soybeans and for USD/CHF.
Let us emphasize that we are not testing the significance of the mean returns, but the returns discounted by a forecast for the volatility.
Roughly, this corresponds to a trading strategy with increasing positions (and less cash) for low volatility forecasts, and decreasing positions (and more cash) for high volatility forecasts, in effect having a portfolio at approximately constant volatility.

\begin{figure}
	\centering
	\includegraphics[width=1\linewidth]{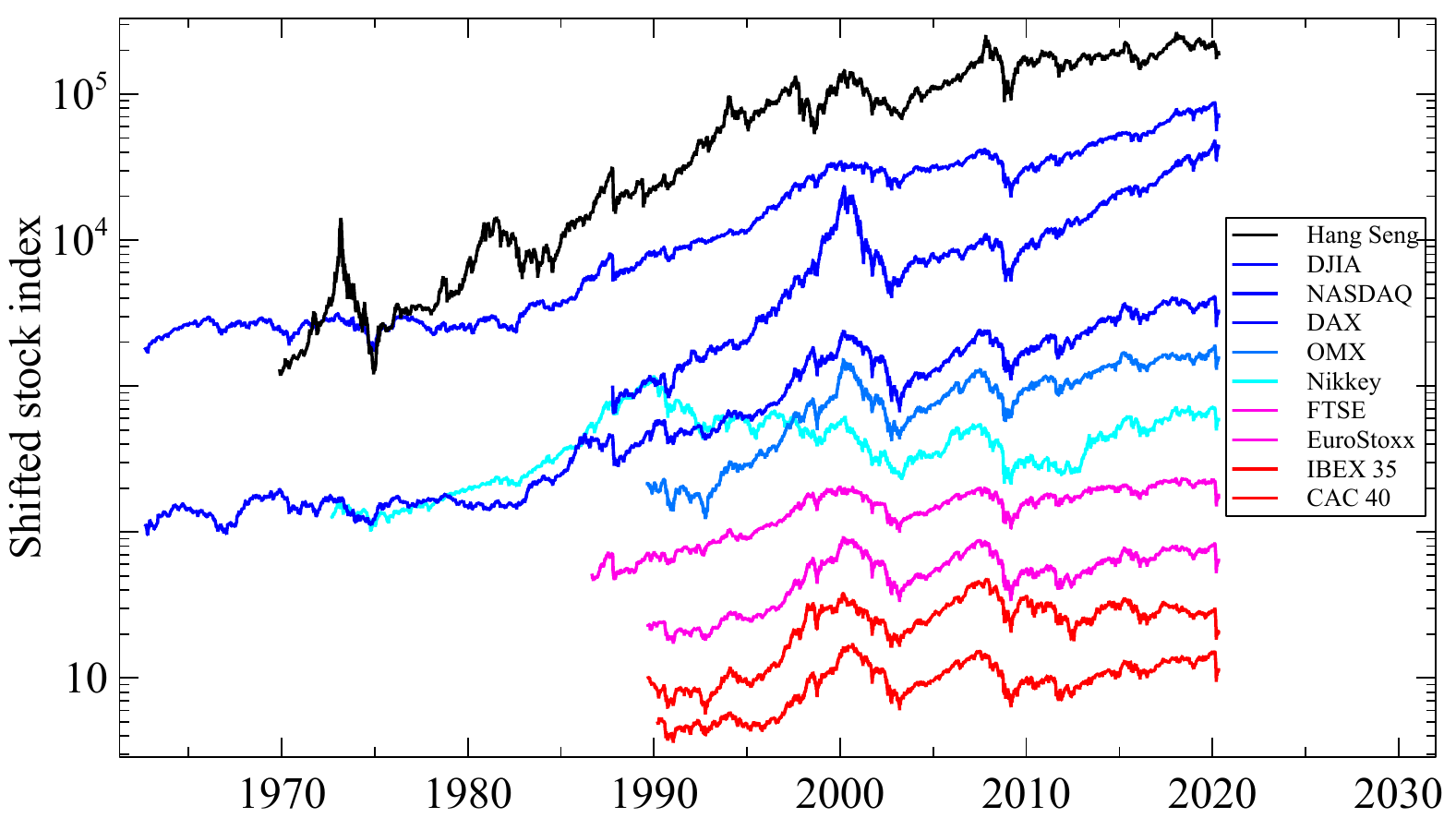}
	\caption{
		The value of the stock index used to compute the significance of the mean innovations.
		The curves are ordered from bottom to top by increasing significance, with the curves in red as not-significant, the ones in blue and black as significant. The labels are ordered accordingly. See the table~\ref{table:LMARCH} for the p-values.}
	\label{fig:stockindexes}
\end{figure}
The stock indexes is the most interesting group.
Fig.~\ref{fig:stockindexes} shows the time series values on a logarithmic scale, with arbitrary shifts set so that the significance of a non-zero mean is increasing for curves higher in the graph.
The interpretation of the significance values is not straightforward, because the economy is an evolving system, and because the time series lengths are not equal (between $\sim$30 to $\sim$50 years).
Therefore, we have a mix between the influence of $n$ (sample size) and the period (particular sample).
Interestingly, over the last 30 years, some indexes do not show a significant deviation from zero.
This should be a caveat for investment advisers on the common statement that the stock market is a good investment over the long term.
Notice that they are market weighted indexes, but not total return indexes, namely they do not include the possible dividends.
For the indexes with the largest $p$-values (curves at the bottom of graph \ref{fig:stockindexes}), an explanation is that the possible increases of values has been distributed to the stock holders as dividends.
With this interpretation, the large $p$-values for the mean indicate that the underlying companies are more redistributing the increased yields than capitalizing for the future.
Interestingly, only European indices are in this case.

For the commodities, only the soybeans index moved significantly, from the range 200 to 350 between 1973 to 2008, then in the range 350 to 600 since 2008.
The other commodity indices are not significantly departing from zero, including precious metal, energy, coffee or chocolate.
With a $p$-value of 8\%, gold is close to have a significant upward move.
Its evolution shows values in the 200 to 500 range up to 2010, and in the 600 to 1100 range after 2010.

In the FX space, only the USD/CHF moved significantly, from roughly 1 USD for 3 CHF in the '70 to an approximate parity since 2010.
This is a factor 3 appreciation of the CHF, in effect during 35 years (the last decade is fluctuating around parity).
All other currencies did not move significantly against the USD.
Let us emphasize that several decades of daily data have been used to reach these results, and it not possible to conclude when using only a few years of data.

\section{Empirical results for the variances}
\label{sec:empiricalVariance}
We now turn to the analysis of the variance of the daily innovations $\epsilon$ computed with LM-ARCH weights, using the statistics on  $\widehat{\sigma^2}$.
For all time series, the variance is larger than 1, and the departure is significant.
All the variances are between 1.05 and 1.09, without obvious difference between asset classes.
The outlier is the USD/CHF exchange rate, which has a much larger variance of 1.17.
This is due to the floor removal by the Swiss National Bank (SNB) of the minimum rate of 1.20 CHF per EUR on January 15, 2015.
The SNB gave no prior indication to the market about the forthcoming change of policy, and on this day, the innovation for USD/CHF is a staggering 44.5!
This event is arguably the largest shocks on a single asset for the last decade.
Since this innovation value is entering the variance with a square, this single day explains the much larger empirical value.

With the hypothesis of a normal distribution for the innovations, the departure from 1 is always significant, with a $p$-value smaller than $10^{-3}$ in all cases (values not reported in the table).
With a Student distribution with 6 degrees of freedom, the results are always significant at a 5\% level, and significant at the 1\% levels except for 3 stock indexes.
With 5 degrees of freedom, all values are significantly above 1 at the 5\% level, except the CAC 40 and the Nasdaq.
Let us emphasize that the theoretical variance of $\epsilon^2$ for a Student distribution has a $1/(\nu - 4)$ singularity (see Eq.~\ref{eq:y_variance.Student}), hence the last distribution is close to the theoretical singularity.
Clearly, assuming a distribution with larger tails makes more difficult to have significant $p$-values.
Let us emphasize that decades of data are required to show significant deviations from unity for a Student distribution.

The departure from 1 is always on the high side, namely toward an explosive grow.
A departure of 5\% for the variance corresponds to an expected doubling of the volatility in 28 business days!
Such an exponential volatility grow is clearly not observed in the financial market, at least on long enough sample.

The difference between the model assumptions and the empirical values points to an incomplete model.
Because on the high side, the empirical time series tend to be more reactive, with occasional large innovations.
This behavior indicates a system that adapt quickly to changing conditions, with a possible over reaction.
As emphasized in Sec.~\ref{sec:archTheory}, the corresponding theoretical model is unstable, but it is sufficient to add a small mean reverting term to stabilize it.
For example, an Ornstein-Uhlenbeck term can be added in the return equation
\begin{equation}
r(t+\dt) = -\,\frac{p(t) - p(t-n)}{\sqrt{n}\,p(t-n)} \,\frac{1}{\tau} + \tilde{\mu}(t) + (1+\gamma) \,\widetilde{\sigma}(t)\, \epsilon(t+\dt)
\end{equation}
where $\gamma \geq 0$ controls the short term instability in the model (with the same conditions as above on $\widetilde{\sigma}$ and the distribution for $\epsilon$).
The parameter $\tau$ controls the intensity of the pull-back term, and is expressed in day (an implicit $\dt$ = 1 day time step is assumed in the equation).
The parameter $n$ controls the lag for measuring price changes, and is also a free parameter.
This term introduces a mean reversion toward past prices and limits the price changes over a longer time interval, hence limits the volatility excursions.
The validation of such processes against empirical data will be difficult, because the deviations from the current process are small.

In the recent literature on stochastic volatility, a new thread concerns the rough volatility models (see \citep{GatheralJaissonRosenbaum} for the original paper).
The overall idea and construction of this model is quite different from the ARCH processes, and involves an independent process for the volatility witch uses a fractional Brownian motions (with Hurst exponent $H < 1/2$) in order to describe the volatility dynamics.
An interesting empirical point is that the realized volatility (defined using high frequency data) shows short term fluctuations of the volatility larger than the usual Brownian motion.
This feature leads to the name ``rough volatility''.
Then, the negative auto-correlations introduced by the fractional Brownian motion in the volatility increments avoid too large excursions of the volatility.

The ARCH models are conceptually different since the volatility is computed from the past price path, and is not an independent hidden process.
The present work points to a fine shortcoming of the ARCH models, in a similar direction as the rough volatility model, namely to a short term roughness of financial time series and to a mean reversion for the volatility.

\section{Empirical results with other volatility estimators}
\label{sec:otherVolatilityEstimators}

\begin{table}
	\begin{tabular}{|lrr|rr|rrr|}
		\hline
		&   &  & \multicolumn{2}{|c|}{mean}  &  \multicolumn{3}{|c|}{variance}  \\
		&   &  &  &  &   &  \multicolumn{2}{c|}{p-value}  \\
		name &  startDate &  length &  mean &  p-value &  variance &  St 6 &  St 5 \\
		\hline
		CAC 40                              & 1989-06-08 & 7862  & 0.016 & 0.156 & 1.187 & 0 & 0 \\
		DAX                                 & 1961-12-04 & 14717 & 0.025 & 0.003 & 1.162 & 0 & 0 \\
		Dow Jones Industrial Average        & 1961-12-01 & 14773 & 0.030 & 0     & 1.209 & 0 & 0 \\
		EuroStoxx                           & 1988-11-30 & 8094  & 0.024 & 0.031 & 1.187 & 0 & 0 \\
		FTSE 100                            & 1985-11-29 & 8736  & 0.024 & 0.025 & 1.215 & 0 & 0 \\
		Hong Kong Hang Seng Index           & 1969-11-24 & 12407 & 0.042 & 0     & 1.217 & 0 & 0 \\
		IBEX 35                             & 1988-12-05 & 7946  & 0.018 & 0.112 & 1.209 & 0 & 0 \\
		NASDAQ 100                          & 1987-01-05 & 8428  & 0.050 & 0     & 1.221 & 0 & 0 \\
		Nikkei 225                          & 1971-12-06 & 11942 & 0.029 & 0.002 & 1.183 & 0 & 0 \\
		OMX STOCKHOLM 30 INDEX              & 1988-11-17 & 7922  & 0.031 & 0.006 & 1.177 & 0 & 0 \\
		\hline
		S\&P GSCI Agriculture               & 1971-12-01 & 12240 & 0.015 & 0.089 & 1.107 & 0 & 0 \\
		S\&P GSCI Coffee                    & 1982-12-08 & 9468  & 0.014 & 0.180 & 1.129 & 0 & 0 \\
		S\&P GSCI Cotton                    & 1978-12-08 & 10476 & 0.004 & 0.709 & 1.137 & 0 & 0 \\
		S\&P GSCI Energy                    & 1984-12-07 & 8962  & 0.006 & 0.578 & 1.266 & 0 & 0 \\
		S\&P GSCI Gold                      & 1979-12-07 & 10224 & 0.019 & 0.049 & 1.114 & 0 & 0 \\
		S\&P GSCI Grains                    & 1971-12-01 & 12241 & 0.017 & 0.065 & 1.106 & 0 & 0 \\
		S\&P GSCI Light Energy              & 1971-12-03 & 12239 & 0.014 & 0.125 & 1.139 & 0 & 0 \\
		S\&P GSCI Non Precious Metals       & 1975-12-09 & 11229 & 0.006 & 0.529 & 1.161 & 0 & 0 \\
		S\&P GSCI Soybeans                  & 1971-12-01 & 12241 & 0.02 & 0.026 & 1.125 & 0 & 0 \\
		\hline
		CAD-USD                             & 1972-12-04 & 12397 &-0.014 & 0.116 & 1.123 & 0 & 0 \\
		CHF-USD                             & 1972-12-04 & 12397 & 0.025 & 0.005 & 1.169 & 0 & 0 \\
		EUR-USD                             & 1976-12-02 & 11175 & 0.001 & 0.892 & 1.123 & 0 & 0 \\
		GBP-USD                             & 1972-12-04 & 12391 &-0.009 & 0.343 & 1.106 & 0 & 0 \\
		JPY-USD                             & 1972-12-04 & 12397 & 0.025 & 0.005 & 1.146 & 0 & 0 \\
		\hline
	\end{tabular}
	\caption{Statistics for $\widehat{\mu}$ and $\widehat{\sigma^2}$ with their p-values.
		The volatility is conputed using equal weights in a window of 500 points.}
	\label{table:HistoricalReturns}
\end{table}

\begin{table}
	\begin{tabular}{|lrr|rr|rrr|}
		\hline
		&   &  & \multicolumn{2}{|c|}{mean}  &  \multicolumn{3}{|c|}{variance}  \\
		&   &  &  &  &   &  \multicolumn{2}{c|}{p-value}  \\
		name &  startDate &  length &  mean &  p-value &  variance &  St 6 &  St 5 \\
		\hline
		CAC 40                              & 1990-03-15 & 7630  & 0.012 & 0.303 & 1.137 & 0 & 0 \\
		DAX                                 & 1962-09-10 & 14494 & 0.027 & 0.001 & 1.147 & 0 & 0 \\
		Dow Jones Industrial Average        & 1962-09-07 & 14546 & 0.036 & 0     & 1.150 & 0 & 0 \\
		EuroStoxx                           & 1989-09-06 & 7859  & 0.017 & 0.127 & 1.188 & 0 & 0 \\
		FTSE 100                            & 1986-09-05 & 8505  & 0.021 & 0.056 & 1.144 & 0 & 0 \\
		Hong Kong Hang Seng Index           & 1969-11-24 & 12373 & 0.050 & 0     & 1.195 & 0 & 0 \\
		IBEX 35                             & 1989-09-11 & 7721  & 0.012 & 0.301 & 1.190 & 0 & 0 \\
		NASDAQ 100                          & 1987-10-12 & 8199  & 0.051 & 0     & 1.142 & 0 & 0 \\
		Nikkei 225                          & 1972-09-11 & 11718 & 0.023 & 0.013 & 1.190 & 0 & 0 \\
		OMX STOCKHOLM 30 INDEX              & 1989-08-24 & 7697  & 0.030 & 0.008 & 1.154 & 0 & 0 \\
		\hline
		S\&P GSCI Agriculture               & 1972-09-06 & 12014 & 0.012 & 0.183 & 1.125 & 0 & 0 \\
		S\&P GSCI Coffee                    & 1983-09-14 & 9241  & 0.015 & 0.136 & 1.154 & 0 & 0 \\
		S\&P GSCI Cotton                    & 1979-09-14 & 10250 & 0.012 & 0.224 & 1.146 & 0 & 0 \\
		S\&P GSCI Energy                    & 1985-09-13 & 8736  & 0.013 & 0.233 & 1.148 & 0 & 0 \\
		S\&P GSCI Gold                      & 1980-09-12 & 9999  & 0.019 & 0.061 & 1.197 & 0 & 0 \\
		S\&P GSCI Grains                    & 1972-09-06 & 12015 & 0.012 & 0.177 & 1.127 & 0 & 0 \\
		S\&P GSCI Light Energy              & 1972-09-08 & 12013 & 0.014 & 0.125 & 1.122 & 0 & 0 \\
		S\&P GSCI Non Precious Metals       & 1976-09-14 & 11004 & 0.010 & 0.282 & 1.137 & 0 & 0 \\
		S\&P GSCI Soybeans                  & 1972-09-06 & 12015 & 0.026 & 0.004 & 1.132 & 0 & 0 \\
		\hline
		CAD-USD                             & 1973-09-10 & 12164 &-0.015 & 0.096 & 1.120 & 0 & 0 \\
		CHF-USD                             & 1973-09-10 & 12164 & 0.022 & 0.013 & 1.252 & 0 & 0 \\
		EUR-USD                             & 1977-09-08 & 10950 & 0.002 & 0.842 & 1.134 & 0 & 0 \\
		GBP-USD                             & 1973-09-10 & 12158 &-0.002 & 0.800 & 1.238 & 0 & 0 \\
		JPY-USD                             & 1973-09-10 & 12164 & 0.012 & 0.182 & 1.209 & 0 & 0 \\
		\hline
	\end{tabular}
	\caption{Statistics for $\widehat{\mu}$ and $\widehat{\sigma^2}$ with their p-values.
		The volatility is computed using EMA weights with factor 0.94.}
	\label{table:EMA}
\end{table}

Other shapes for the volatility kernel can be used.
The results for a rectangular moving average (RMA) of length 500 and an exponential moving average (EMA) with decay factor 0.94 are reported in this section.
The sample lengths vary slightly, due to the different initialization lengths of the different kernels.
This can impact mostly the statistics $\widehat{\mu}$, to a lesser extent $\widehat{\sigma^2}$.
For the mean values, the figures are essentially consistent with the LM-ARCH kernel.
The largest systematic differences occur for $\widehat{\sigma^2}$, with typical values in the range 1.1 to 1.2, to be compared with the range 1.05 to 1.09 with a LM-ARCH kernel.
Due to these larger statistics, all the p-values are less than $10^{-3}$, namely the value 1 is always excluded with a high significance, including with a distributional assumption of a Student with 5 degrees of freedom.

The higher values for the variances show that these alternative kernels are less good at whitening the returns, and the empirical data are further away from the related theoretical processes.
Another hypothesis of the process is that the innovations should be i.i.d., namely independent and identically distributed.
Independence is not tested using only the variance, but a new powerful ``tile test'' has been presented recently \citep{Zumbach.TileTest} that focuses on the time dependency in a forecast setting for risk evaluations.
The results of the tile test are consistent with the present ones, namely the LM-ARCH kernel is excellent at transforming empirical returns into innovations that are close to the theoretical hypothesis.
The previous section shows that the LM-ARCH is very good, but not perfect.

\section{Further investigations}
\label{sec:furtherInvestigations}
We have tried some variations around the present LM-ARCH kernel, changing the shape of the decay and/or its finite length.
Although we have certainly not tested all possibilities, no clear improvements can be achieved, and most changes produce systematically higher values for the empirical variances.
Hence, it seems difficult to reach unit empirical variances by modifying only the shape of the volatility kernel.

Can a unit variance be recovered by using an affine variance estimator instead of a linear one?
Using the process set-up as described in Sec.~\ref{sec:archTheory}, the answer is no.
The equations have been set such that
$\E{\widetilde{\sigma^2}_\text{aff}} = \E{\widetilde{\sigma^2}_\text{lin}} = \E{r^2} = \sigma^2_\infty$, regardless of $w_\infty$.
The parameter $\sigma^2_\infty$ fixes all the mean variance, and there is no parameter left with that respect.
Hence, in order to get more freedom in the model, some equations have to be modified.

One possibility is to not use the equation \ref{eq:locSizeShape} for the process, while still using \ref{def:historicalInnovation} to define the innovations.
This corresponds to a purely empirical set-up, where the data are generated by some unknown process, and another variable called innovation is defined from the historical returns.
The relation between linear and affine variance is still given by Eq.~\ref{eq:AffVarianceFromLinVariance}, and the affine variance is used to discount the historical return.
In order to recover innovations with a unit variance, we impose the further condition
\begin{eqnarray*}
\E{\widetilde{\sigma^2}_\text{aff}} = (1+\gamma) \,\E{\widetilde{\sigma^2}_\text{lin}}
\end{eqnarray*}
Taking the expectation of Eq.~\ref{eq:AffVarianceFromLinVariance} and combining with this relation, we obtain
\begin{equation*}
  \sigma^2_\infty = \E{r^2} \left( 1 + \frac{\gamma}{w_\infty} \right).
\end{equation*}
Using plausible values for the parameters, $\gamma$ = 0.07 and $w_\infty$ = 0.035, the mean variance parameter should be 3 times larger than the historical variance of the data.
This large factor originates in the small value for $w_\infty$, which should be compensated by a large $\sigma^2_\infty$ in order to have an impact on the final variance.
Clearly, this does not make sense to have such a discrepancy between the empirical sample variance and the corresponding parameter, and deeper modifications of the equations should be introduced.

In the ARCH model, we have decided to neglect the mean return term $\widetilde{\mu}(t)$, because difficult to estimate and dominated by the volatility.
We can do a similar hypothesis for the innovations, and measure the size of the innovations with
\begin{align}
\frac{1}{n-1}\,\sum_t \left(\epsilon(t)\right)^2.
\end{align}
Overall, the results using this estimator show only minor modifications compared to the estimator \eqref{eq:empiricalVariance}, and in particular the significance of the departure from 1 are almost identical.

The empirical study can be extended to longer time interval for the innovations.
The idea is to compute an \textit{ex-ante} volatility forecast for a horizon $\DT$, to get the \text{ex-post} realized returns at the corresponding horizon $\DT$, and finally to compute the empirical innovations at $\DT$.
With increasing time interval $\DT$, the innovation variances are typically growing, the effective sample sizes $n_\text{efffective} = n/\DT$ are decreasing, and the threshold values for statistical significance are also growing.
Beside the present computations with $\DT$ = 1 day, we have done the investigation for $\DT$ = 2, 5, 10 and 30 days.
Overall, the innovation variances is growing at the same pace as the significance threshold, but the dispersion between the time series is increasing.
The statistical results for the above empirical sample are thus less clear cut, with the variances for a few time series that are not significantly above 1.
Nevertheless, the majority of innovation variances are significantly above 1 (for a Student distribution with 6 degrees of freedom and at a 5\% level).

\section{Conclusions}
\label{sec:conclusion}
The family of ARCH processes is very good at describing financial time series, and can capture most of known stylized facts.
The model used in this study can be extended in several ways, for example to include the leverage effect presents for stocks and stock indexes.
Yet, its very simplicity and robustness is an important advantage when dealing with large number of time series, and for various asset types, in particular for risk estimations.
In this context, the LM-ARCH process proves to be very efficient (see \citep{Zumbach.TileTest} for a recent back-test of risk evaluations).
Moreover, the theoretical backing by a quadratic ARCH process allows deriving consistent volatility forecasts at increasing risk horizons.

The volatility forecast is a crucial ingredient in a risk computation, the following one is the distribution for the innovations.
The usual assumption is that the variance for the innovations should be 1, for all time series.
This study shows that the empirical variances are slightly but significantly above one.
In order to reach this conclusion, and in particular to have significant values at a 5\% level, empirical samples of 3 to 5 decades are necessary.
The empirical variances are in a range from 1.05 to 1.09, a small but significant deviation from the theoretical process with a unit value.

The empirical variances are always above 1, indicating a systematic short term instability and a potentially explosive grow for the volatility.
This points to a fast adaptability of the financial system to a changing environment, with the trading activity and the induced prices that react quickly to changing conditions.
On the theoretical side, this point to an incomplete model, with a possible short term instability that must be compensated by a stabilization term acting at intermediate time horizons.
Since this is a small deviation from the current model, it will be difficult to validate such an extension.

Another theoretical condition is that the mean of the innovations should be zero.
Most commodity indexes and FX are compatible with this hypothesis, apart from the soybean index and USD/CHF exchange rate.
For stock indexes, it is widely believed that they are rising on the long term, hence a positive mean innovation is expected to be significant.
Overall, this is the case, except clearly for the CAC 40 and IBEX 35, and marginally for the EuroStoxx and FTSE 100.
These zero expected innovations mean that the possible increase in the underlying stock prices have been distributed in form of dividends (or that there was no significant increase of values!).

\bibliographystyle{plainnat}
\bibliography{bibliography}

\end{document}